\documentclass[reprint,prl,amsmath,amssymb,floatfix,aps]{revtex4-2}
\usepackage{graphicx}
\usepackage{bm}
\usepackage{amsmath}
\usepackage{amssymb}
\usepackage{datetime}
\begin{document}
\author{Tigran V. Shahbazyan}
\email{shahbazyan@jsums.edu}
\affiliation{Department of Physics, Jackson State University, Jackson, MS 39217, USA}
\title{Purcell factor for plasmon-enhanced metal photoluminescence}


\begin{abstract}
We present an analytical model for the plasmonic enhancement of metal photoluminescence (MPL) in metal nanostructures with a characteristic size below the diffraction limit. In such systems, the primary mechanism of MPL enhancement is the excitation of localized surface plasmons (LSP) by recombining carriers followed by photon emission due to LSP radiative decay. For plasmonic nanostructures of \textit{arbitrary} shape, we obtain a universal expression for the MPL Purcell factor that describes the plasmonic enhancement of MPL in terms of the metal dielectric function,  LSP frequency, and system volume. We find that the lineshape of the MPL spectrum is affected by the interference between  direct carrier recombination processes and those mediated by plasmonic antenna which leads to a blueshift of MPL spectral band  relative to LSP resonance in scattering spectra observed in numerous experiments.
\end{abstract}

\maketitle

\clearpage
\section{Introduction}

The photoluminescence of noble metals \cite{mooradian-prl69,apell-ps88,boyd-prb86} 
and  metal nanostructures  \cite{boyd-prb86,wilcoxon-jcp98,novotny-prb03,feldmann-prb04,fourkas-nl05,eichelbaum-nanotech07,biagioni-prb09,loumaigne-nl10,wissert-nl10,hulst-nl11,biagioni-nl12,elsayed-cpl00,elsayed-jpcb03,elsayed-prb05,bouhelier-prl05,imura-jpcb05,imura-jpcc09,wang-oe09,link-jpcc11,orrit-nl12,link-acsnano12,shen-acsnano12,meixner-jpcc13,meixner-jpcc13-2,wu-acsnano15,lu-jpcc19,wang-nl07,potma-jpcc08,schuck-prl05,link-acsnano15,klar-nl16,lei-acsnano17,klar-nl18} 
has attracted continuing interest fueled, to  a large extent, by its  numerous applications, e.g., in imaging \cite{wang-pnas05,durr-nl07,park-oe08,chen-acsnano10,orrit-biophysj16,shigemoto-advmat09}, 
sensing \cite{wackenhut-jpcc21,hupp-nl01,krenn-bionanosci11,gong-jpcc12,gong-jpcc12},
nanothermometry \cite{kall-acsphot18,baffou-acsnano21,orrit-nl18},
and optical recording \cite{zijlstra-nature09}. 
In bulk metals, the underlying mechanism of metal photoluminescence (MPL) is the radiative recombination of photoexcited d-band holes  and upper-energy sp-band electrons  via a momentum-conserving interband transition, a process that competes against much faster nonradiative transitions in noble metals, such as Auger scattering, resulting in extremely low MPL quantum yields\cite{mooradian-prl69,apell-ps88,boyd-prb86} $\sim 10^{-10}$. 
Much brighter MPL has been reported from various metal nanostructures supporting localized surface plasmons (LSP), including nanospheres \cite{wilcoxon-jcp98,novotny-prb03,feldmann-prb04,fourkas-nl05,eichelbaum-nanotech07,biagioni-prb09,loumaigne-nl10,wissert-nl10,hulst-nl11,biagioni-nl12}, 
nanorods \cite{elsayed-cpl00,elsayed-jpcb03,elsayed-prb05,bouhelier-prl05,imura-jpcb05,imura-jpcc09,wang-oe09,link-jpcc11,orrit-nl12,link-acsnano12,shen-acsnano12,meixner-jpcc13,meixner-jpcc13-2,wu-acsnano15,lu-jpcc19}, 
nanowires \cite{wang-nl07,potma-jpcc08}, nanoparticle dimers \cite{schuck-prl05,link-acsnano15,klar-nl16,lei-acsnano17}, 
and porous structures \cite{klar-nl18}. In all such systems, the generic MPL spectrum represents a nearly Lorentzian peak centered close to  the LSP frequency  with a peak amplitude depending weakly on the system size  \cite{novotny-prb03,feldmann-prb04,fourkas-nl05,shen-acsnano12}, except for very small systems, in which  MPL is largely quenched  \cite{feldmann-prb04,shahbazyan-nl13}, or for large structures, in which LSP is radiatively  damped \cite{novotny-prb03,fourkas-nl05,shen-acsnano12}.
 
The origin of bright MPL from plasmonic  structures was suggested \cite{feldmann-prb04} to be due to the excitation of the LSP by a recombining electron-hole pair \cite{shahbazyan-prl98,shahbazyan-cp00,shahbazyan-nl13} followed by the LSP radiative decay. This plasmonic antenna scenario has  been extensively tested and largely accepted based mainly on MPL measurements from single gold nanorods  which revealed strong similarities between the MPL spectra and  scattering spectra measured from the same structures \cite{link-jpcc11,orrit-nl12,link-acsnano12,shen-acsnano12,meixner-jpcc13,meixner-jpcc13-2,wu-acsnano15,lu-jpcc19}. More recent experiments on hot carrier recombination in plasmonic  structures have highlighted the important role of intraband transitions, but here as well,  strong similarities between the MPL and scattering spectra point to  the dominant role of  LSP-mediated processes \cite{baumberg-nl15,cahill-pnas14,baumberg-nl17,ren-acsnano16,link-acsnano18,link-acsnano20,link-jcp21}. At the same time,  persistent differences in the lineshape and in peak positions between the MPL and  scattering spectra have  been widely reported in diverse nanostructures under various excitation conditions \cite{link-acsnano12,orrit-nl12,wu-acsnano15,lei-acsnano17,ren-acsnano16,link-acsnano18,link-jcp21}. Notably, the MPL spectra are  blueshifted relative to scattering spectra, which has not so far been explained.

The enhancement of the radiation rate is described by the Purcell factor \cite{novotny-book}, which, for an emitter situated near a plasmonic structure, is determined by the LSP's local density of states (LDOS) at the emitter's position \cite{shahbazyan-prl16,shahbazyan-prb18}. In contrast, the MPL Purcell factor describes the enhancement of the time-averaged signal and, therefore, involves spatial averaging of the LDOS over the metal volume, which requires the calculation of local fields  within the entire structure. Such calculations, with the rare exception of high symmetry systems,  present a considerable numerical challenge \cite{link-acsnano18}. On the other hand, the surprising similarity of MPL spectra across diverse systems
\cite{boyd-prb86,wilcoxon-jcp98,novotny-prb03,feldmann-prb04,fourkas-nl05,eichelbaum-nanotech07,biagioni-prb09,loumaigne-nl10,wissert-nl10,hulst-nl11,biagioni-nl12,elsayed-cpl00,elsayed-jpcb03,elsayed-prb05,bouhelier-prl05,imura-jpcb05,imura-jpcc09,wang-oe09,link-jpcc11,orrit-nl12,link-acsnano12,shen-acsnano12,meixner-jpcc13,meixner-jpcc13-2,wu-acsnano15,lu-jpcc19,wang-nl07,potma-jpcc08,schuck-prl05,link-acsnano15,klar-nl16,lei-acsnano17,klar-nl18} 
suggests that a simpler, yet accurate, description of plasmon-enhanced MPL should be possible within an analytical approach.

In this article, we present an analytical model for MPL from plasmonic structures of \textit{arbitrary} geometry whose characteristic size is below the diffraction limit. For such systems, we derive an explicit expression for the MPL Purcell factor, which describes the plasmonic antenna effect, and trace the observed blueshift of the MPL spectral band to destructive  interference between the direct and antenna-assisted recombination processes. For frequencies $\omega$ close to  the LSP frequency $\omega_{n}$, we obtain the MPL enhancement factor relative to the bulk MPL background in the following universal form
\begin{equation}
\label{spectrum-mpl}
M_{n}(\omega)
=A_{n}\left |\frac{\varepsilon'(\omega_{n})-1}{\varepsilon(\omega)-\varepsilon'(\omega_{n})}\right |^{2}, 
\end{equation}
where $\epsilon(\omega)=\epsilon'(\omega)+i\epsilon''(\omega)$ is  the metal dielectric function 
and the parameter $A_{n}$  weakly depends  on the system volume.  For \textit{any} system geometry, the lineshape of the MPL spectrum  is determined solely by the metal dielectric function and  LSP frequency.  For larger structures, Eq.~(\ref{spectrum-mpl}) is extended to include the LSP radiation damping effect while retaining its universal form.

\section{MPL Purcell Factor}

We start by recalling that MPL involves several stages \cite{boyd-prb86,apell-ps88,link-acsnano18} including the photoexcitation of the nonequilibrium population of d-band and sp-band carriers, carrier energy and momentum relaxation due to electron-electron and electron-phonon scattering, and carrier recombination accompanied by photon emission, all of which contribute to the MPL overall quantum yield. Here, we focus on the latter process which can involve LSP excitation in the intermediate state \cite{shahbazyan-prl98}. Note that only a small fraction of recombination processes, for which the emitted photon energies fit within the LSP spectral band, undergo plasmonic enhancement \cite{feldmann-prb04}. For interband transitions, which require no momentum relaxation mechanism,  the transition matrix element has the form $\int dV_{\rm m}\psi_{k}^{sp}\Phi\psi_{k}^{d}\approx -\bm{d}_{k}\!\cdot\!\bm{E} (\bm{r})$, where $\psi_{k}^{sp}$ and $\psi_{k}^{d}$ are the Bloch wave functions for the sp-band and d-band, respectively,  $\bm{d}_{k}$ is the interband dipole matrix element between electron states with the same wave vector $\bm{k}$, $\Phi(\bm{r})$ is the local potential and $\bm{E} (\bm{r})= - \bm{\nabla}\Phi(\bm{r})$ is the corresponding  local field at  d-band hole's position $\bm{r}$, while the integral is taken over the metal volume $V_{\rm m}$ \cite{shahbazyan-nl13}. Note that, in contrast to the radiation field that is nearly uniform on the system scale, the local field $\bm{E} (\bm{r})$ can vary significantly  inside the nanostructure,  so a recombination of a low-mobility d-band hole and   sp-band electron can be viewed as a dipole transition that takes place at some position $\bm{r}$ inside the nanostructure. 

The rate at which the energy of the excited electron-hole pair is transferred to an electromagnetic (EM) environment is \cite{novotny-book} $\Gamma_{k}=(2/\hbar) \text{Im} \left [\bm{d}_{k} \bm{D}(\omega;\bm{r},\bm{r})\bm{d}_{k}\right ]$, 
where $\bm{D}(\omega;\bm{r},\bm{r}')$ is  the EM dyadic Green function obeying (in operator form) $\bm{\nabla}\times\bm{\nabla}\times\bm{D} - (\omega^{2}\varepsilon/c^{2})\bm{D}=(4\pi \omega^{2}/c^{2}) \bm{I}$; here $\varepsilon(\omega,\bm{r})$ is the system dielectric function equal to $\varepsilon(\omega)$ in the metallic regions  and $\varepsilon_{d}$ in the  dielectric ones (we set $\varepsilon_{d}=1$ for now),  $c$ is the speed of light, and $\bm{I}$ is the unit tensor. To extract the LSP contribution to $\bm{D}$, we note that, for unretarded electron motion,  the LSP modes are described by the longitudinal fields $\bm{E}_{n} (\bm{r})= - \bm{\nabla}\Phi_{n}(\bm{r})$ and obey the lossless Gauss's equation \cite{stockman-review} $\bm{\nabla}\cdot[\varepsilon'(\omega_{n},\bm{r})\bm{\nabla}\Phi_{n}(\bm{r}) ]=0$. 
The dyadic Green function $\bm{D}$ relates to the scalar Green function  $D$ for the potentials $\Phi$ as $\bm{D}(\omega;\bm{r},\bm{r}')=\bm{\nabla}\bm{\nabla}'D(\omega;\bm{r},\bm{r}')$, where $D$ satisfies $\bm{\nabla}\cdot\left [\varepsilon(\omega,\bm{r}) \bm{\nabla} D (\omega;\bm{r},\bm{r}')\right ]=4\pi \delta (\bm{r}-\bm{r}')$. We now split the scalar Green function as  $D=D_{0}+D_{\rm LSP}$, where $D_{0}=-|\bm{r}-\bm{r}'|^{-1}$ is the free-space near-field Green function and $D_{\rm LSP}$ is  the LSP contribution.  Expanding the latter
over the LSP modes as $D_{\rm LSP} (\omega;\bm{r},\bm{r}')=\sum_{n}D_{n}\Phi_{n}(\bm{r})\Phi_{n}(\bm{r}')$, we obtain the expansion coefficients as
\begin{equation}
\label{mode-coeff}
D_{n}(\omega)= -  \dfrac{4\pi}{\int\! dV \varepsilon (\omega,\bm{r})\bm{E}_{n}^{2}(\bm{r})},
\end{equation}
where we omitted a constant term \cite{shahbazyan-prb18}. Since,  due to   Gauss's law, we have $\int\! dV \varepsilon'(\omega_{n},\bm{r})\bm{E}^{2}_{n}(\bm{r})=0$, the expansion coefficients $D_{n}(\omega)$ exhibit LSP poles in the complex frequency plane at $\omega=\omega_{n}-i\gamma_{n}/2$, where $\gamma_{n}=2\varepsilon''(\omega_{n})/[\partial \varepsilon'(\omega_{n})/\partial \omega_{n}]$ is the LSP decay rate (we assume $\gamma_{n}/\omega_{n}\ll 1$). Presenting $D_{n}(\omega)$ as the sum over LSP poles would lead to  normal field expansion suitable for quantum approaches \cite{shahbazyan-prb21}. However, within classical approaches, it is suitable to obtain another representation for the LSP Green function (\ref{mode-coeff}) directly in terms of the metal dielectric function. Namely, the denominator is presented as $\int\! dV \varepsilon (\omega,\bm{r})\bm{E}_{n}^{2}(\bm{r})=
\left [\varepsilon (\omega)-\varepsilon' (\omega_{n})\right ] \int\! dV_{\rm m}\bm{E}_{n}^{2}(\bm{r})$,
where the integral is taken over the metallic regions while contributions from the dielectric regions, characterized by constant permittivities, cancel out. We thus obtain the LSP Green function for the electric fields as
\begin{equation}
\label{plasmon-green2}
\bm{D}_{\rm LSP} (\omega;\bm{r},\bm{r}')=\sum_{n}\dfrac{-4\pi}{\int\! dV_{\rm m} \bm{E}_{n}^{2}}\frac{\bm{E}_{n}(\bm{r})\bm{E}_{n}(\bm{r}')}{\varepsilon (\omega)-\varepsilon' (\omega_{n})}.
\end{equation}
Note that even though $\bm{E}_{n}(\bm{r})$ represents eigenmodes of the lossless Gauss equation, the Green function (\ref{plasmon-green2}) is valid for the complex dielectric function $\varepsilon (\omega)=\varepsilon'(\omega)+i\varepsilon''(\omega)$ as well. Indeed, if $\epsilon''$ is included as a perturbation, then the first-order correction leads to Eq.~(\ref{mode-coeff}), and the higher-order corrections involve nondiagonal terms of the form $\varepsilon''(\omega)\!\int \! dV_{\rm m}\bm{E}_{n}(\bm{r}) \cdot \bm{E}_{n'}(\bm{r})$, which vanish due to the modes orthogonality \cite{shahbazyan-prb21}, so that the LSP Green function (\ref{plasmon-green2}) is, in fact, exact in all orders. The energy transfer rate from the emitter to the plasmonic environment takes the form
\begin{equation}
\label{rate}
\Gamma_{k}(\omega,\bm{r})=\dfrac{2}{\hbar}\sum_{n}\dfrac{ [\bm{d}_{k} \!\cdot \!\bm{E}_{n}(\bm{r})]^{2}}{\int\! dV_{\rm m} \bm{E}_{n}^{2}}\,\text{Im}\!\left [\frac{-4\pi}{\varepsilon (\omega)-\varepsilon'(\omega_{n})}\right ].
\end{equation}
Note that the rate (\ref{rate}) is suitable for an emitter situated  either outside or inside the plasmonic structure, but here we  consider only the latter case.

We now turn to the MPL Purcell factor, which is  defined as $F_{n}=\langle\Gamma_{k}\rangle/\gamma_{k}^{\rm r}$, where $\gamma_{k}^{\rm r}=4d_{k}^{2}\omega^{3}/3c^{3}\hbar$ is the emitter's radiative recombination rate \cite{novotny-book} and  $\langle\Gamma_{k}\rangle$ stands for the average of rate (\ref{rate}) over the emitter's  positions  inside the metal and over orientations of its dipole moment $\bm{d}_{k}=d_{k}\bm{n}_{k}$. For relatively small nanostructures under long illumination, the electron-hole pairs are distributed nearly uniformly. Using $\langle \bm{n}_{k} \bm{n}_{k} \rangle =\frac{1}{3}\bm {I}$ and averaging Eq.~(\ref{rate}) over the metal volume, we obtain the MPL Purcell factor \textit{per} emitter in the following \textit{universal} form,
\begin{equation}
\label{purcell2}
F_{n}(\omega)
=\dfrac{2\pi c^{3}}{\omega^{3}V_{\rm m}}
\frac{\varepsilon'' (\omega)}{\left |\varepsilon (\omega)-\varepsilon'(\omega_{n})\right |^{2}},
\end{equation}
where we kept only the resonant term. The MPL Purcell factor is inversely proportional to the metal volume, while its frequency dependence, apart from  normalization, is determined by the metal dielectric function. At resonance  frequency $\omega=\omega_{n}$, the maximal value of the Purcell factor is $F_{n}=2\pi c^{3}/\omega_{n}^{3}V_{\rm m}\varepsilon'' (\omega_{n})$, which is related to the LSP mode volume ${\cal V}_{n}$ as \cite{shahbazyan-acsphot17,shahbazyan-nl19} $F_{n}=6\pi c^{3}Q_{n}/\omega_{n}^{3}{\cal V}_{n}$, where $Q_{n}=\omega_{n}/\gamma_{n}$ is the LSP quality factor, and we obtain
\begin{equation}
\label{mpl-mode-volume}
{\cal V}_{n}=3Q_{n}\varepsilon''(\omega_{n})V_{\rm m}=\dfrac{3}{2}V_{\rm m}\,\omega_{n}\,
\partial \varepsilon'(\omega_{n})/\partial \omega_{n}.
\end{equation}
The LSP mode volume (\ref{mpl-mode-volume}) for MPL can be significantly larger than the metal volume $V_{\rm m}$, especially for long-wavelength LSPs characterized by large $|\varepsilon'(\omega_{n})|$. Note, however, that the LSP mode volume plays \textit{no} significant role in MPL, in contrast to that for an emitter situated  outside the structure \cite{shahbazyan-prl16,shahbazyan-prb18}, as it effectively cancels out in the radiated power spectrum (see  below).

\section{Emission Spectrum}

We now turn to the emission spectrum of a recombining electron-hole pair mediated by the excitation of LSP. For systems with a characteristic size below the diffraction limit, the radiated power spectrum is \cite{novotny-book} $W_{k}^{\rm r}(\omega)=(\omega^{4}/3c^{3})|\bm{p}_{k}(\omega)|^{2}$, where $\bm{p}_{k}(\omega)=\bm{p}_{k}^{\rm eh}(\omega)+\bm{p}_{k}^{\rm pl}(\omega)$ is the system dipole moment  composed of the electron-hole pair's dipole moment $\bm{p}_{k}^{\rm eh}(\omega)$ and the induced dipole moment of the plasmonic structure $\bm{p}_{k}^{\rm pl}(\omega)$. To determine $\bm{p}_{k}^{\rm eh}(\omega)$, we note that, due to the very fast relaxation rate of the d-band hole,  the recombination events are  not correlated. In the linear regime, the interband polarization $\rho_{k}$ can be determined from the Maxwell-Bloch equations
\begin{equation}
\label{mb}
i\partial \rho_{k}/\partial t=(\omega_{k}-i\gamma_{k}/2)\rho_{k}-\bm{d}_{k}\!\cdot\!\bm{E} (\bm{r})/\hbar
\end{equation}
with the  initial condition $\rho_{k}(0)=1$. Here, $\omega_{k}$ is the interband transition frequency, $\gamma_{k}$ is the intrinsic decay rate due to the  fast Auger processes \cite{apell-ps88}, and $\bm{E} (\bm{r})$ is the local field at the emitter position. The coupling to LSP is included in the standard way \cite{novotny-book}, by relating the local field back to the interband dipole moment  $\bm{p}_{k}^{\rm eh}=\bm{d}_{k}\rho_{k}$ as $\bm{E} (\bm{r})=\bm{D}_{\rm LSP}(\bm{r},\bm{r})\bm{p}_{k}^{\rm eh}$. Upon performing the Laplace transform, the interband dipole moment takes the form
\begin{equation}
\label{dipole-pair}
\bm{p}_{k}^{\rm eh}=\dfrac{\bm{d}_{k}}{\omega_{k}-\omega -i\gamma_{k}/2-i\Gamma_{k}/2},
\end{equation}
where the energy transfer rate  $\Gamma_{k}$ is given by Eq.~(\ref{rate}). 

The MPL enhancement comes from the induced dipole moment of plasmonic structure defined as
%
\begin{equation}
\bm{p}_{k}^{\rm pl}\!=\!\!\int\! dV\chi(\omega,\bm{r}) \bm{E}(\bm{r})\!=\!\!\int\! dV'\chi(\omega,\bm{r}') \bm{D}_{\rm LSP}(\omega;\bm{r}',\bm{r})\bm{p}_{k}^{\rm eh},
\end{equation}
where $\chi=(\varepsilon-1)/4\pi$ is the system susceptibility. With help of the LSP Green function (\ref{plasmon-green2}), we obtain
\begin{equation}
\bm{p}_{k}^{\rm pl}= -\sum_{n}\dfrac{4\pi\bm{p}_{n}(\omega)}{\int\! dV_{\rm m} \bm{E}_{n}^{2}}\frac{\bm{E}_{n}(\bm{r}) \cdot\bm{p}_{k}^{\rm eh}}{\varepsilon (\omega)-\varepsilon'(\omega_{n})},
\end{equation}
where $\bm{p}_{n}(\omega)=\int dV\chi(\omega,\bm{r}) \bm{E}_{n}(\bm{r})$ is the dipole moment of the LSP mode. Finally, using Eq.~(\ref{dipole-pair}), the radiated power spectrum  for an electron-hole pair takes the form
\begin{equation}
\label{emission2}
W_{k}^{\rm r}(\omega)
=\dfrac{\omega^{4}}{3c^{3}|\Omega_{k}|^{2}}
\left |\bm{d}_{k}
-\sum_{n}\dfrac{4\pi\bm{p}_{n}(\omega)}{\int\! dV_{\rm m} \bm{E}_{n}^{2}}
\frac{\bm{E}_{n}(\bm{r})\cdot\bm{d}_{k}}{\varepsilon (\omega)-\varepsilon'(\omega_{n})}\right |^{2},
\end{equation}
where $\Omega_{k}=\omega_{k}-\omega -i\gamma_{k}/2-i\Gamma_{k}/2$.  

The full radiated power spectrum $W^{\rm r}(\omega)$ is obtained by the $\bm{k}$-integration of Eq.~(\ref{emission2})  with the factor $f_{k}^{\rm sp}(1-f_{k}^{\rm d})$, where $f_{k}^{\rm sp}$ and $f_{k}^{\rm d}$ are the Fermi distribution functions for the sp-band and d-band, respectively. After orientational averaging, we obtain  $W^{\rm r}=W_{\rm eh}^{\rm r}M$, where
%
\begin{equation}
\label{direct}
W_{\rm eh}^{\rm r}=\frac{\omega^{4}}{3c^{3}}\sum_{\bm k}\frac{d_{k}^{2}f_{k}^{\rm sp}(1-f_{k}^{\rm d})}{|\omega_{k}-\omega -i\gamma_{k}/2-i\Gamma_{k}/2|^{2}}
\end{equation}
describes direct emission by the recombining carriers and 
\begin{equation}
\label{enh}
M(\omega,\bm{r})=\left \langle \left |\bm{n}_{k}
-\sum_{n}\dfrac{4\pi\bm{p}_{n}(\omega)}{\int\! dV_{\rm m} \bm{E}_{n}^{2}}
\frac{\bm{E}_{n}(\bm{r})\cdot\bm{n}_{k}}{\varepsilon (\omega)-\varepsilon'(\omega_{n})}\right |^{2}\right \rangle_{\!\bm{n}_{k}}
\end{equation}
is the local MPL enhancement factor. The direct emission spectrum (\ref{direct}) differs from the bulk MPL spectrum by the presence of an additional relaxation rate $\Gamma_{k}$, which describes the excitation of high-momentum LSPs by recombining carriers and, for small systems, leads to  MPL quenching \cite{shahbazyan-nl13}. Note  that due to the extremely short d-band hole lifetime ($<50$ fs), MPL is significantly quenched only for very small  ($\lesssim 3$ nm) structures, in which most excited carriers are  sufficiently close to the metal surface \cite{feldmann-prb04,shahbazyan-nl13}. Otherwise, we have 
$W_{\rm eh}^{\rm r}=qW_{\rm b}^{\rm r}$, where $W_{\rm b}^{\rm r}$ is the bulk MPL spectrum and $q\lesssim 1$ is the quenching parameter  that weakly depends on the system volume.

Turning now to the MPL enhancement factor (\ref{enh}), the main contribution comes from the  LSP (second) term. Keeping only the resonance term (with $\omega$ close to $\omega_{n}$) and neglecting, for now, the direct (first) term, the LSP contribution to the MPL spectrum takes the form
\begin{equation}
\label{enh-pl}
M_{n}^{\rm LSP}(\omega,\bm{r})
=\dfrac{1}{3}
\left [\dfrac{\bm{E}_{n}(\bm{r})}{\int\! dV_{\rm m} \bm{E}_{n}^{2}}\right ]^{2}
\left |\frac{4\pi\bm{p}_{n}(\omega)}{\varepsilon (\omega)-\varepsilon'(\omega_{n})}\right |^{2},
\end{equation}
where the factor of 1/3 comes from orientational averaging. Noting that $\bm{p}_{n}(\omega)=\chi(\omega)\int\! dV_{\rm m}\bm{E}_{n}$ and averaging Eq.~(\ref{enh-pl}) over the metal volume, we obtain
\begin{equation}
\label{spectrum-pl}
M_{n}^{\rm LSP}(\omega)
=\dfrac{s_{n}}{3}\left |\frac{\varepsilon(\omega)-1}{\varepsilon (\omega)-\varepsilon'(\omega_{n})}\right |^{2}, 
\end{equation}
where the parameter 
\begin{equation}
\label{sp}
s_{n}=\dfrac{\left (\int\! dV_{\rm m}\bm{E}_{n}\right )^{2}}{V_{\rm m}\int\! dV_{\rm m} \bm{E}_{n}^{2}}
\end{equation}
is  independent of the field overall amplitude and, for the  dipole LSP modes, of the system volume as well. 
For nanospheres and nanospheroids, its exact value is $s_{n}=1$, while smaller values are expected for other geometries.
To pinpoint the exact mechanism of MPL  enhancement, we note that, for small nanostructures, the LSP's radiation efficiency is  $\eta_{n}=w_{n}^{\rm r}/w_{n}^{\rm nr}$, where $w_{n}^{\rm r}$ and $w_{n}^{\rm nr}$ are its radiated power and the Ohmic losses, respectively \cite{novotny-book}:
\begin{equation}
w_{n}^{\rm r}(\omega)=\dfrac{\omega^{4}|\bm{p}_{n}(\omega)|^{2}}{3c^{3}},
~~
w_{n}^{\rm nr}(\omega)=\dfrac{\omega\varepsilon'' (\omega)}{8\pi}\!\int\! dV_{\rm m}\bm{E}_{n}^{2}(\bm{r}).
\end{equation}
Then the LSP contribution  (\ref{spectrum-pl}) can be recast as 
\begin{equation}
M_{n}^{\rm LSP}(\omega)=\eta_{n}(\omega)F_{n}(\omega),
\end{equation}
where the Purcell factor $F_{n}(\omega)$ is given by Eq.~(\ref{purcell2}), implying that MPL enhancement is indeed due to the plasmonic antenna effect. Since, for small nanostructures, the antenna's radiation efficiency is proportional to its  volume, the latter effectively cancels out the Purcell factor's volume dependence, as reflected in the parameter $s_{n}$.

We now turn to the effect of interference between the direct and LSP-mediated MPL processes, described by the cross term in Eq.~(\ref{enh}). After performing  spatial and orientational averaging, we obtain 
\begin{equation}
\label{spectrum-int}
M_{n}^{\rm int}(\omega)
=-\dfrac{2s_{n}}{3}\text{Re}\left [\frac{\varepsilon(\omega)-1}{\varepsilon (\omega)-\varepsilon'(\omega_{n})}\right ]. 
\end{equation}
Combining $M_{n}^{\rm LSP}(\omega)$ and $M_{n}^{\rm int}(\omega)$  and omitting a constant term, we arrive at the enhancement factor  $M_{n}(\omega)$, given by Eq.~(\ref{spectrum-mpl}), with $A_{n}=s_{n}q/3$ that includes the quenching parameter $q$.  Comparing the frequency dependence of $M_{n}(\omega)$  and $M_{n}^{\rm LSP}(\omega)$, we observe that $\varepsilon(\omega)$ in the numerator of Eq.~(\ref{spectrum-pl}) is  replaced with $\varepsilon'(\omega_{n})$ in Eq.~(\ref{spectrum-mpl}), implying that interference between  the direct and LSP-mediated   transitions, however weak, can substantially affect the MPL spectral band (see below). 

The enhancement factor (\ref{spectrum-mpl}) is derived for small  structures by assuming that the LSP's radiation damping  is much weaker than the Ohmic losses in the metal. For larger systems, the radiation damping effect can be included in the standard way by considering the system's interaction with the radiation field \cite{carminati-oc06,novotny-book}. The result reads
\begin{equation}
\label{enh-mpl-rad}
M(\omega)
=\sum_{n}A_{n}\left |\frac{\varepsilon'(\omega_{n})-\varepsilon_{d}}
{\varepsilon(\omega)-\varepsilon'(\omega_{n})-\frac{2i}{3}k^{3}V_{n}[\varepsilon(\omega)-\varepsilon_{d}] }\right |^{2}, 
\end{equation}
where we restored the environment permittivity $\varepsilon_{d}$. Here, $k=\sqrt{\varepsilon_{d}}\,\omega/c$ is the light wave vector, $V_{n}=V_{\rm m}|\varepsilon'(\omega_{n})/\varepsilon_{d}-1|s_{n}/4\pi$ is the effective volume \cite{shahbazyan-pra23}, and the sum runs over  LSP modes that couple to the far field (e.g., longitudinal and transverse dipole modes in nanorods). For the nanosphere of radius $a$, we have $s_{n}=1$ and $\varepsilon'(\omega_{n})=-2\varepsilon_{d}$, and we recover  $V_{n} =a^{3}$. The enhancement factor (\ref{enh-mpl-rad})  can be used to describe  the MPL spectra for systems of various shapes and sizes with no substantial numerical effort. It is also suitable for describing MPL spectra of complex or random metallic structures, such as nanosponges \cite{klar-nl18}, which exhibit multiple LSP peaks. In this case, the coefficients $A_{n}$ can be tuned to fit the  spectral peaks' amplitudes.

\section{Numerical results}

To illustrate our results, below we present the MPL spectra for several gold nanostructures immersed in water ($\varepsilon_{d}=1.77$) by plotting Eq.~(\ref{enh-mpl-rad}) for a single LSP mode. In all calculations, we use the experimental gold dielectric function. Accordingly, the LSP wavelength range extends above  530 nm, which corresponds to the interband transition onset in gold.  Specifically, we choose the LSP peak positions at  wavelengths  of 620, 670, 710, and 750 nm, which are close to those measured in the MPL experiments\cite{orrit-nl12,link-acsnano12,wu-acsnano15}. We assume no particular nanostructure shape but specify the overall volume as $V_{m}=L^{3}$, where $L$ is the characteristic linear size. For gold structures, the optimal $L$ lies in the interval between 20  and 60 nm, in which the MPL quenching is relatively weak while the LSP radiative damping  is not too strong \cite{novotny-prb03,feldmann-prb04,fourkas-nl05,shen-acsnano12,shahbazyan-nl13}, so we set $q=1$, $s_{n}=1$, and $A_{n}=1/3$.   

\begin{figure}[tb]
  \begin{center}
\includegraphics[width=0.99\columnwidth]{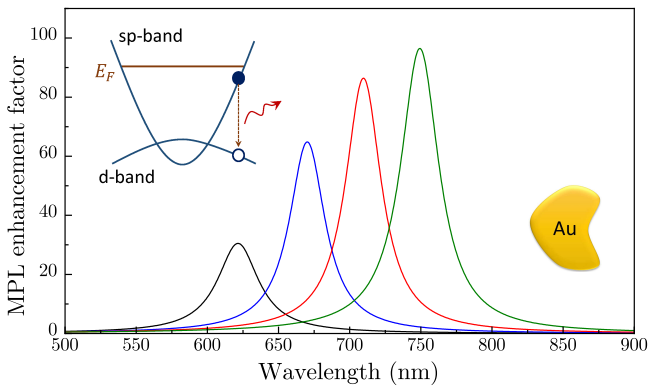}
  \end{center}
\caption{\label{fig1} MPL enhancement factor for general-shape gold  nanostructures with characteristic size 20 nm at various LSP wavelengths of 620, 670, 710, and 750 nm. (Inset) Schematic of interband transitions in gold.}
  \end{figure}

In Fig.~\ref{fig1}, we plot the MPL enhancement factor $M_{n}(\omega)$  for general-shape gold  structures  with  $L=20$ nm, in which the LSP resonance position can be tuned by varying the structure shape (e.g., by varying the aspect ratio of the nanorods). The MPL enhancement factor exhibits nearly-Lorenzian resonances, whose amplitude increases with the wavelength, consistent with  plasmon-enhanced MPL quantum yield measurements \cite{link-acsnano12,orrit-nl12}. Note that the bulk MPL exhibits the opposite trend, as its efficiency is greater for shorter wavelengths due to an increased sp-band electron density of states for higher energies close to the Fermi level \cite{klar-nl16}.

\begin{figure}[tb]
  \begin{center}
\includegraphics[width=0.99\columnwidth]{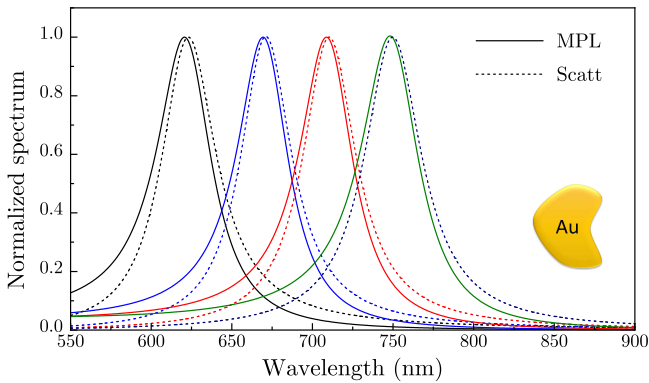}
  \end{center}
\caption{\label{fig2} Normalized MPL and scattering spectra for gold nanostructures with a characteristic size of 40 nm at the same LSP wavelengths as in Fig.~\ref{fig1}. }
\end{figure}

In Fig.~\ref{fig2}, we compare the MPL spectra for  $L=40$ nm gold nanostructures at the same LSP wavelengths obtained as $S_{\rm MPL}(\omega)\propto \omega^{4}M_{n}(\omega)$  and the scattering spectra, which are related to the plasmonic antenna's emission as $S_{\rm scatt}(\omega)\propto \omega^{4}M_{n}^{\rm LSP}(\omega)$, both  normalized to their maxima. 
A clear blueshift of the MPL spectral band relative to the LSP resonance in  scattering spectra persists for all structures but  is more pronounced for shorter wavelengths, consistent with the experiment \cite{link-acsnano12,orrit-nl12,wu-acsnano15,lei-acsnano17,ren-acsnano16,link-acsnano18,link-jcp21}. Specifically, the calculated normalized spectra closely resemble those measured for gold nanostructures of various shape\cite{link-acsnano12,wu-acsnano15}. This blueshift and the change in the resonance lineshape originate from the interference between the direct and LSP-mediated recombination processes, as discussed above, resulting in the replacement $\varepsilon(\omega)\rightarrow\varepsilon'(\omega_{n})$ in the numerator of Eq.~(\ref{enh-mpl-rad}). 

Finally,  our approach should apply to the \textit{intraband} MPL as well, where similar differences between the MPL and scattering spectra were reported \cite{ren-acsnano16,link-acsnano18,link-jcp21}. Although any intraband MPL mechanism must include additional momentum relaxation channels, the final-state transition should involve both the direct and LSP-mediated processes,  so the intraband MPL spectrum should still be described by Eq.~(\ref{enh-mpl-rad}), albeit with different $A_{n}$, which incorporates the interference between them. Note that the calculated normalized spectra in Fig. \ref{fig2} closely resemble the intraband MPL spectra recorded in silver nanorods\cite{ren-acsnano16}.

\section{Conclusions}

We have developed an analytical approach to plasmon-enhanced MPL that accurately describes the emission spectra of metal nanostructures of arbitrary geometry with the characteristic size below the diffraction limit. We have established that the primary mechanism of MPL enhancement is excitation of LSPs by recombining carriers followed by the LSP radiative decay, and we have obtained an explicit universal expression for the MPL Purcell factor in terms of the metal dielectric function, LSP frequency, and  system volume. We have shown that the precise lineshape of the MPL spectrum is defined by the interference between the direct and LSP-mediated photon emission processes, resulting in the blueshift of the MPL spectra relative to scattering spectra reported in numerous experiments. 
Our results can be used for modeling   experimental MPL spectra for small nanostructures of arbitrary shape without extensive numerical effort.

\acknowledgements
This work was supported in part by  National Science Foundation grants  DMR-2000170,  DMR-1856515,  and  DMR-1826886.


{}

\end{document}